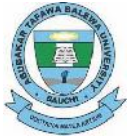

# An Extensive Survey of Digital Image Steganography: State of the Art

Idakwo M. A., Muazu M. B., Adedokun A. E., Sadiq B. O.
*Department of Computer Engineering,*
*Faculty of Engineering,*
*Ahmadu Bello University, Zaria Kaduna State Nigeria.*

**ABSTRACT**
*The need to protect sensitive information privacy during information exchange over the internet/intranet has led to wider adoption of cryptography and steganography. The cryptography approaches convert the information into an unreadable format however draws the attention of cryptanalyst owing to the uncommon random nature flow of the bytes when viewing the flowing structured bytes on a computer. While steganography, in contrast, conceals the very existence of covert communication using digital media. Although any digital media (text, image, video, audio) can covey the sensitive information, the media with higher redundant bits are more favorable for embedding the sensitive information without distorting the media. Digital images are majorly used in conveying sensitive information compared to others owing to their higher rate of tolerating distortions, highly available, smaller sizes with high redundant bits. However, the need for maximizing the redundancy bits for the optimum embedding of secret information has been a paramount issue due to the imperceptibility prerequisite which deteriorates with an increase in payload thus, resulting in a tradeoff. This has limited steganography to only applications with lower payload requirements, thus limiting the adoption for wider deployment. This paper critically analyzes the current steganographic techniques, recent trends, and challenges.*

**INTRODUCTION**

Protecting and guarantying long-distance communication security has been a critical issue in information exchange. It becomes more paramount if sensitive information is to be conveyed on the internet which is open and public in nature as a result of a higher increase in cybercrimes activities (Yahya, 2019). Sequel to ensuring secure communication, variance of cryptographic techniques suchlike Rivest Shamir and Adleman (RSA), international data encryption algorithm (IDEA), advance encryption standard (AES), data encryption standard (DES), blowfish, among others were used to obscure sensitive information as they scramble the information into an unreadable and unintelligent format referred to as ciphertext (Elhoseny *et al.*, 2018). Such ciphertexts are only transformed into a readable format with the right cryptography key. Unfortunately, they raise curiosity and draw cryptanalyst attention. This increases attacks on the ciphertexts using brute force

*Corresponding author: Idakwo, M. A.* ✉ *mondabutu@gmail.com* ✉ *Department of Computer Engineering, Faculty of Engineering, Ahmadu Bello University, Zaria.* © *2020. Faculty of Technology Education, ATBU Bauchi. All rights reserved*

(Zaheer *et al.*, 2019). Therefore, the desire to hide the communicating medium for the purpose of avoiding eavesdroppers' attention using steganography (Xu & Nie, 2018). In order to evade eavesdropper attention, digital steganography was introduced. Steganography is a Greek word formed by the combination of two words steganous which means "covered" and graphy which means "writing". Therefore, it means covered writing (Atawneh *et al.*, 2017). Steganography is the art and science of concealing communication within other information (image, text, audio, video, etc.) (Emad *et al.*, 2018).

Regardless of all the efforts and huge successes been achieved in steganography, the adoption rate of steganography is still very low in real-time applications where privacy of sensitive information is paramount such as health records, military reports, forensic reports amongst others (Kadhim *et al.*, 2020). Rather, cryptography is seeing as a good alternative even thou it arouses suspicion. The reason for this, hinges on the fact that information contents vary in volume especially in health records, military reports, amongst others**.** It is therefore imperative that for any steganography system to withstand such large information volume, there must be an equivalent larger capacity referred to as payload to accommodate the information in a way that image quality is maintained known as imperceptibility, equally a secure stego-key to restrict unauthorized users with the ability to withstand any image processing techniques that can alter or change the embedded information is important. (Yahya, 2019)**.** Perchance, these issues have lingered on and remain a perpetual issue for researchers as a consequence of the strict constraint of imperceptibility on the steganographic system. The quest to address these issues has resulted in trading off a better balance between imperceptibility and payload. This paper critically reviews the recent achievements in digital image steganography including their requirements, operations, evaluation metrics in the spatial domain, transform domain and adaptive domain.

**DIGITAL IMAGE STEGANOGRAPHY**

A computer sees an image as numbers in array forms that represents light intensities at various locations referred to as pixels (Sathua *et al.*, 2017). The pixels form the images raster data and determine the file size. An image with 640 × 480 pixels and 256 colors (i.e 8 bits in each pixel) for example, may accommodate 300 kilobits of data (Penha *et al.*, 2017). There are usually stored either as 24-bit or 8-bit format, nevertheless, the 24-bit images have the highest capacity for concealing sensitive information as a pixel can be represented by 16,777,216 color values. The images with 24-bit are usually larger except for images in JPEG formats (Muhammad *et al.*, 2017). All pixels color variations are obtained from blue, red and green which are the primary colors and are represented as a byte each. Every image has redundant bits that can be altered with information without distorting the image quality and it is referred to as digital image steganography.

The sensitive information is concealed in this redundant bits of image referred to as cover-image using an algorithm and most time with a key known as stego-key to produce an output image known as stego-image. The general steganographic system model is as shown in Figure 1.

*Corresponding author: Idakwo, M. A. ✉ mondabutu@gmail.com ✉ Department of Computer Engineering, Faculty of Engineering, Ahmadu Bello University, Zaria.* 

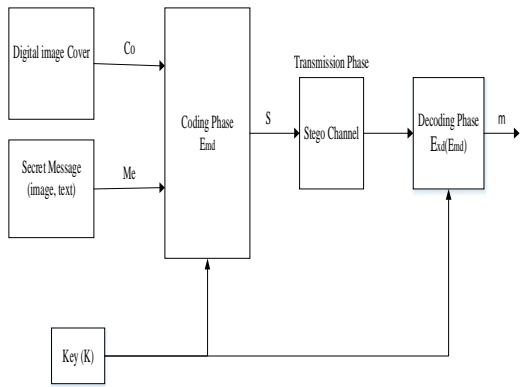

**Figure 1:** Digital Steganographic System Model

The embedding schemes entail the sensitive information, the stego-key controlling the extraction and embedding algorithm, with the cover-image used in transmitting the sensitive information. The extraction algorithm performs a reverse of the embedding operations and retrieves sensitive information. The embedding and extracting algorithms are represented by (Yahya, 2018):

$$Ed : C_i \oplus K \oplus M_s \rightarrow S_i \quad (1)$$

$$Ex : S_i \oplus K \rightarrow M_s \quad (2)$$

Such that;

$$Ex(Ed(c.k.m)) \approx m, \forall c \in C_i, k \in K, m \in M_s$$

Where $Ed$ is the embedding function, $Ex$ is the extracting function, $C_i$ is the cover-image, $K$ is the stego-key, $M_s$ is the sensitive information, $S_i$ is the stego-image to be communicated.

The crucial factors used to characterize the performances of the steganographic system are imperceptibility and embedding payload (Yahya, 2019). Where imperceptibility indicates the difficulties by human eyes in distinguishing between a generated stego-image from the obtained cover-image, while the total capacity an image can carry is referred to as embedding payload (Atawneh *et al.*, 2017).

***Steganalysis overview***

Unlike cryptography, it is impossible to identify a stego-image by monitoring the channel used for communicating. Equally, the several steganographic embedding processes has hardened the decoding of a stego-image without any knowledge of the embedding processes. This independence nature of steganographic methods easily nullifies the current methods of decoding (Qi *et al.*, 2016). Thus, statistical analysis techniques are basically the sole method for detecting steganography (Douglas *et al.*, 2018). Contrarily, steganography is the steganalysis, which continually strives to detect hidden communication. It is mainly focused on obtaining the statistical anomaly in the stego-image caused during the information concealing process. Therefore, steganalysis remains a critical threat to steganography (Atawneh *et al.*, 2017). Intentionally, programs suchlike first or second-order statistics, correlation within pixels, distances and directions histograms, amongst others can be written to evaluate and compare the cover0image and stego-image statistical properties (Florez *et al.*, 2018). In steganalysis, all efforts aiming to destroy traces of hidden communication without revealing the sensitive information are known as passive steganalysis (Yahya, 2018). Several current steganalysis methods are only effective on some types of image steganographic techniques whose steganography methods are known in advance (Douglas *et al.*, 2018). Practically, it is infeasible in an internet environment where billions of images are been transmitter, it becomes difficult to determine if an image is carrying information or not (Qi *et al.*, 2016). So, current steganalysis techniques are only used in analyzing a group of suspected images instead of monitoring the whole internet

*Corresponding author: Idakwo, M. A. ✉ mondabutu@gmail.com ✉ Department of Computer Engineering, Faculty of Engineering, Ahmadu Bello University, Zaria.* *© 2020. Faculty of Technology Education, ATBU Bauchi. All rights reserved*

image exchange. Therefore, the analysis is only applicable to existing known steganography methods (Qi *et al.*, 2016). Furthermore, either knowingly or unknowingly, image processing techniques (image transformation, addition of random noise, rotations and scaling, cropping or decimation, compression) can altered the embedded sensitive information. The resistance of a stego image to all the image processing techniques is refer to as robustness (Yahya, 2019).

#### STEGANOGRAPHIC TECHNIQUES

Using the embedding processes as a criterion, the digital image is classified into spatial or time, frequency or transform and adaptive domain (Laishram & Tuithung, 2018).

### *Spatial Domain*

The spatial domain suchlike least significant bit (LSB) (Lou & Hu, 2012), pallet base pixel value differencing (PVD) (Ma *et al.*, 2015), and diamond encoding (DE) (Kuo *et al.*, 2015) are based on direct modification of pixel intensities and have larger embedding capability with slight degradation of image quality (Hussain *et al.*, 2018). Although, subjecting the stego-image to simple or hybrid attacks such as cropping, compression, rotation, Gaussian noise with salt and pepper, and translation will destroy the embedded information, hence, there are not robust against attacks (Muhammad *et al.*, 2017). Various time-domain stenography information concealing techniques have been proposed in literature suchlike exploiting modification direction (EMD), LSB matching, LSB amongst others (Jindal & Kaur, 2016). The LSB technique developed in 1989 by Turner in 1989 is highly use owning to their ease in replacing the $p^{th}$ bits $(1 \leq p \leq 8)$ of a pixel with the sensitive message. The generated stego-image will maintain good quality once $p \leq 3$.

Unfortunately, any steganographic analytical techniques can detect them easily (Lee *et al.*, 2017). Therefore, the LSB matching became a substitute and it avoids directly modifications of bits with the secret information, reduced the numbers of pixels that can be modified and equally withstand chi-square attacks. The LSB matching technique conceals inside two bits in pixels and modifies only a bit. bits modify each pixel by one bit and maintain a quality image compared to LSB (Liao *et al.*, 2017). In other to enhance the numbers of bits that can be modified, the exploiting modification direction (EMD) was introduced. There are varying EMD methods and they both uses their neighboring pixels in embedding the sensitive information.

### *Transform Domain*

The frequency or transform domain suchlike discrete cosine transform (DCT), discrete Fourier transform (DFT), discrete wavelet transform (DWT), integer wavelet transform (IWT) amongst others methods decomposes the image into frequency coefficients before embedding sensitive information in them making them robust against attacks and imperceptible. A system is said to be robust if it has the capabilities to withstand all attacks that can modify or alter the concealed information and the system abilities to withstand distortion is referred to as imperceptible (Atawneh *et al.*, 2017). Nonetheless, the transform domain has a lower payload with a higher computational complexity which makes them slow (Kaur & Pandey, 2017).

In steganography, the discrete Fourier techniques use a lot of Fourier information during the reconstruction of the local signal which makes it a poor tool in reconstructing a non-smooth signal.

Therefore, the DCT technique became a good alternative but required selecting the DCT coefficients carefully to maintain a good quality (Cheddad *et al.*,

*Corresponding author: Idakwo, M. A. ✉ mondabutu@gmail.com ✉ Department of Computer Engineering, Faculty of Engineering, Ahmadu Bello University, Zaria.* *© 2020. Faculty of Technology Education, ATBU Bauchi. All rights reserved*

2010) and is computationally complex with low payload (Vanathi, 2017). The DWT became an alternative as it is flexible and adaptive to the human visual system. Nonetheless, DWT generates coefficients with floating-point values and truncating it into an integer since pixels have integers value that may alter or destroy the embedded information (Arunkumar *et al.*, 2019). This has led to the current adoption of integer wavelet transform which has the ability to map an integer input to integer output to overcome the conventional floating-point issue (Chao & Fisher, 1996; Xiong *et al.*, 2018).

### *Adaptive Domain*

The adaptive steganography equally known as the statistics-aware technique is a hybrid of the spatial domain (Subhedar & Mankar, 2014). This technique evaluates the image statistical global characteristics before attempting to interact with the image frequency coefficients (Hajduk and Levický, 2016). The essence of the statistics is for detecting regions in an image that can be changed without distorting the image. This method is characterized by the selection of a randomly adaptive pixel which relies on the information conveying image and selection of pixels with larger local standard deviation in each block. The selection of pixels in a block based on the standard deviation is for neglecting areas of similar color (e.g. smooth areas in images). This implies that images with complex colors are very good for adaptive steganography. The commonly distinct between time, frequency and adaptive domain are as presented in Table 1

**Table 1:** Comparing time, frequency and Adaptive domain

| Properties | Spatial | Transform | Adaptive |
|---|---|---|---|
| Format | Dependent | Independent | Independent |
| System category | Simple | Complex | Algorithm dependent |
| Pixel Manipulation | Direct | Indirect | Inline technique used |
| Computational Complexity | Less | High | Depends on Algorithm |
| Payload | High | Limited | Varies |
| Imperceptibility | Low | High | Highly controllable |
| Virtual features integrity | Maintainable | Less | Maintainable |
| Robustness | Highly prone | Less | Algorithm dependent |
| Geometric Attacks | Highly vulnearable | Resistant | Highly resistant |
| Statistical detectability (RS, Histogram) | Easy | Hard | Hard |
| Non-structural Attacks | Detectable | detectable | Difficult |
| Focus: Capacity | High | Moderate | Moderate |
| Visual Capacity | High | High | High |
| Undetectability | Moderate | High | High |

*Corresponding author: Idakwo, M. A. ✉ mondabutu@gmail.com ✉ Department of Computer Engineering, Faculty of Engineering, Ahmadu Bello University, Zaria.* *© 2020. Faculty of Technology Education, ATBU Bauchi. All rights reserved*

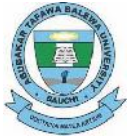

**QUALITATIVE EVALUATION METRIC**

In order to effectively evaluates the quality of stego-images, mean square error (MSE), peak signal-to-noise ratio (PSNR), normalize cross correlation (NCC), structural similarity index (SSIM) and histogram are generally used (Preishuber *et al.*, 2018).

*Mean square error*

This evaluates how accurate a stego-image is by evaluating the variability between the cover-image and stego-image (Atawneh *et al.*, 2017)

$$MSE = \frac{1}{pq}\sum_{m=1}^{p}\sum_{n=1}^{q}(S_{mn} - C_{mn})^2 \quad (3)$$

$p$ and $q$ is the input images number of rows and columns respectively, $C_{mn}$ is the cover-image while $S_{mn}$ is the stego-image

*Peak signal-to-noise ratio (PSNR)*

The PSNR evaluates the rate at which a stego-image is distorted as a result of the embedded information in the unit of decibel (dB). A higher value of PSNR reflects a higher quality of the stego-image and calculated as (Wang *et al.*, 2018)

$$`PSNR = 10\log_{10}\frac{X_{max}^2}{MSE} \quad (4)$$

Where $X_{max}$ is the maximum pixel value in both stego-image and cover-image and is

$$X_{max} \leq \begin{cases} 1 \\ 255 \end{cases} \quad (5)$$

*Structural similarity index (SSIM)*

SSIM compares two given images of similarity using their structures with values ranges between -1 and 1. The closer the SSIM is to 1 the more the similarity between the given two images and it is calculated by (Kumar & Kumar, 2018)

$$SSIM(p,q) = \frac{(2\sim_p \sim_q + k_1)(2\dagger_{pq} + k_2)}{(\sim_p^2 + \sim_q^2 + k_1)(\dagger_p^2 + \sim_q^2 + k_2)} \quad (6)$$

Where the cover-image and stego-image pixels mean intensity are $\sim_p, \sim_q$ respectively, the cover-image and stego-image variance are $\dagger_p, \dagger_q$ and $k_p, k_q$ represent the constant added for stability when $u_p^2 + u_q^2$ it is almost zero. Also $\sim_p = u_q = (kL)^2$, as $L$ is the pixel's dynamic ranges (255 for 8 bit) and $k \ll 1$

*Normalize cross correlation (NCC)*

The NCC evaluates the rate of variation between a stego-image and a cover-image as the either differs or resembles. It reaches its maximum when the two images are similar and it is calculated using equation (Elhoseny *et al.*, 2018).

$$NCC = \frac{n\sum CS - (\sum C)(\sum S)}{\sqrt{n(\sum C^2)-(\sum C)^2}\sqrt{n(\sum S^2)-(\sum S)^2}} \quad (7)$$

Where; $n$ represents the pairs of data number, $C$ represent the cover-image, and $S$ is stego-image

*Bits error rates (BER)*

The BER basically measures the proportion of the original information bits and the distorted information bits due to image attack consequences. It is evaluated mathematically as (Elhoseny *et al.*, 2018).

$$BER = \frac{S_w - S_a}{S_w} \quad (8)$$

Where: $S_w$ is the original stego-image, $S_a$ is the attacked stego-image

*Image histogram*

An image histogram is a graphical method of presenting each pixel's occurrence and distributions. It is used to evaluate stego-image imperceptibility. The histogram measures the extent an embedding process distorted the stego-image. A higher resemblance between the cover-image histogram and stego-image histogram indicated that the distortion was minimal after embedding the secret information into the cover-image (Atawneh *et al.*, 2017).

*Corresponding author: Idakwo, M. A. ✉ mondabutu@gmail.com ✉ Department of Computer Engineering, Faculty of Engineering, Ahmadu Bello University, Zaria.* © *2020. Faculty of Technology Education, ATBU Bauchi. All rights reserved*

**LITERATURE REVIEW**

Several researchers have identified and surveyed the notions, prevalent and current issues in image steganography. The conventional LSB steganography embedding method is susceptible to the chi-square $\left(t^2\right)$ and regular singular attacks. These two well-known steganalysis methods will perfectly estimate any concealed sensitive information length and thus defeating the purpose of secret communication. The regular singular attack detects both randomly and sequentially concealed information with a starting value around $\pm 6\%$. Whereas the $t^2$ detection can only detect embedded information in a sequential form from an initial value of $10\%$. In other to evade these attacks, Lou and Hu (2012) proposed selection rules for cover-images selection and adopted a starting point below $1\%$. Before embedding any stego-key and sensitive information inside the cover-image, each pixel features were assessed and grouped with their image histogram controlled during the embedding process in order to minimize distortion. For full cover-image redundancy bits exploit, Li *et al.* (2013) presented a new data concealing technique using two-dimensional (2D) difference-histogram modification using difference pair mapping (DPM). The DPM is an injective mapping defined in different pairs. All the smooth areas in the pixel that were targeted by the pixel pair selection scheme were precisely located. Hence, all embeddable regions were completely utilized. The cover-image histogram bins were shifted to create embeddable space thus enabling more sensitive information to be hiding in the image's smooth area while maintaining the image quality. Similarly, (Qin *et al.*, 2013) implemented an adaptive reference pixel selection scheme based on the histogram shift and inpainting forecasting model. The techniques select reference pixel using an adaptive scheme based on the image contents and distribution features and predict a similar corresponding image using a partial differential equation. The predicted image has the same structural and geometry information with the original image. Trial and error threshold was used in selecting the reference images and more images were selected images with higher fidelity areas compared to images with smooth areas. In order to enhance imperceptibility while increasing embeddable regions, sensitive information was hidden in the predicted histogram error which lies between two groups of peak and zero points by performing a shift operation. The strategy of selecting embeddable reference pixels based on regions can reduce the distortion effect arising from the embedding method. However, the reference pixel threshold can be optimized using a stochastic optimization algorithm such as particle swamp optimization in order to yield a near-optimal system performance. In order to improve the LSB schemes, Dagar (2014) presented a highly randomized color image steganography with two secret keys (key-1 and Key-2) for information embedding. Key-1 is an array of circular 1D which is limited to either 0 or 1 and Key-2 is an 8 digits 1D array. The key-1 was exclusively OR with the selected colored image most significant bits used in concealing the sensitive information and key-2 were used to decide the concealing locations for the sensitive information. This was necessary as the LSB embedding techniques are easily predictable and susceptible to attacks. The extraction process likewise uses Key1 and Key2 in performing the inverse of the embedding operations. The randomization of the concealing processes with two secret keys improved the system security level as a steganalyst needs the two secret-keys in order to extract the conceal information. However, producing, maintaining and sharing two secret-keys is a tedious and hard task.

Since the conventional thorough pixel row to row raster scan method for embedding secret information is not a prerequisite for optimal PSNR. Kanan and Nazeri (2014) model steganography as an optimization search problem with the aim of obtaining the best starting point and direction for embedding sensitive information in a cover-image while optimizing the stego-image PSNR. A genetic algorithm (GA) was adopted to search the stego-image spaces using PSNR as a fitness function. The GA chromosomes used 7 genes to form the pixel bits and the sensitive bits were converted into their specific genes. A comparison was made between the sensitive and pixels bits before embedding the sensitive information in the pixel's bits. If the sensitive information bits is greater than the pixel's bit, it is an indication that the related chromosome does not have the capacity to embed the sensitive information bits inside the cover-image. Otherwise, the sensitive bits are inserted into the equivalent pixel bit. Similarly, an adaptive transform technique based on LSB-DCT with random embedding capabilities was presented by Habib *et al.* (2015).The secret information was transformed into a 1D binary vector and the cover-image was transformed into 8 by 8 non-overlapping blocks and quantized using 2D DCT to produce the quantized cover-image DCT coefficients. A digital chaotic generator using two perturbed peace wise linear chaotic map designed to produce a binary stream was used in determining the DCT coefficient locations for concealing the sensitive information. Since the LSB has the ability of concealing an information in a sequentially manner, the LSB security was enhanced by using a chaotic generator to randomly embed the information. However, the DCT coefficients are susceptible to high artifacts if not properly selected during the embedding process with lower payload (Cheddad *et al.*, 2010) and huge computational complexity (Vanathi,2017). Therefore, Rabie and Kamel (2016) proposed a solution for imperceptibility and embeddable regions issues by guessing the adaptive areas within a fixed block of DCT in the cover image. The scheme adopted an embedding function and a quantization step to adaptively evaluate the highest region in the SCT block of the cover-image high frequency (lower-right). The sensitive information was straightforwardly inserted inside the square area to replace the cover image DCT coefficients of the exact block regions. These procedures recorded a high embeddable capacity and ensured perceptibility at an acceptable rate. Furthermore, the scheme implemented an optimal cover size automatic selection that suits any given secret information sizes without leaving any redundancy bits, thus minimized transmission bandwidth overhead. The major drawback with the fixed-block scheme adopted is that natural scene images are not stationary statistically over the entire fixed-block support regions. The reason for this is that in the same fixed block, the patterns of brightness are not the same thus varying from one location to the other. Thus, it is important that the concealing techniques suit the block sizes and the cover-image several areas characteristics for maximum embedding capacity. Similar to Rabie and Kamel (2016), Huang and Zhao (2016) derived a new cover selection techniques based on linear error forecasting in time domain steganography. The linear forecasting error model was used in modeling the relationship between the image pixels with the aim of ascertaining how the pixels follows patterns. The forecast error is the measured error while forecasting the accuracy of the detector. For the purpose of establishing the linear forecast error stability and the correlations between linear predictor error and accuracy, images were randomly selected and divided into smaller subsets by the linear forecast error interval using a threshold. All deviations along the linear forecast error were calculated and the corresponding accurate curves were plotted under varying conditions. The obtained results showed that images with higher linear predictor accuracy usually has the least detection accuracy, so therefore, are selected as cover-images. Furthermore, the linear forecast error generated more correlation compared to local variances and entropy in the time domain. The forecast error used in the time domain was able to select a better cover-image with computational time depending on the size of the image database size. More so, natural images pixels are dependent on the adjacent pixels and have limited relevance with distant regions pixels. It is therefore imperative to limit the predictor to the neighboring pixels in order to optimize the predictor time. More so, pixels values in complicated texture areas gain more diversity than those in simple texture areas. Thus, the possibilities of detecting a disturbance or modification in simple texture areas is relatively higher when compared to complicated texture area.

*Corresponding author: Idakwo, M. A. ✉ mondabutu@gmail.com ✉ Department of Computer Engineering, Faculty of Engineering, Ahmadu Bello University, Zaria.* *© 2020. Faculty of Technology Education, ATBU Bauchi. All rights reserved*

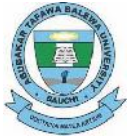

On the other hand, Hajduk and Levický (2016) experimented with a cover image selection steganography using a database of 2000 images. The DWT was adopted in decomposing the images into their coefficients and the signs of the image were removed and formed a preprocessed images datasets. Although the cover selection methods guarantee a quality stego-image since the best cover-image with a large area for a secret bit was chosen. However, decomposing and transforming large thousands of images into their sub bands coefficient increases computational time. As an alternative to using DCT in steganographic applications, the DWT became a good option as it is flexible when representing any image signal with good adaptability to the human visual system (Kumar & Kumar, 2018). Nevertheless, DWT coefficient outputs are in floating-point and truncating them into an integer number may lead to loss of sensitive information as the coefficients of the output wavelet will not be an integer value (Arunkumar *et al.*, 2019) (Raftari & Moghadam, 2012). In order to reduce the distortion encounter in DWT, Atawneh *et al.* (2017) presented an adaptive imperceptible and secured image system using a diamond encoding (DE) in DWT. The DE was used mainly for reducing the artifacts produced in DWT as a result of the floating-point errors and equally to improve the embedding processes. An algorithm was used in converting the sensitive information into a sequence in base-5 digits after which the cover-image was transformed into segmented pair of the coefficient. After this, the DE scheme will now embed the sensitivity into the cover-image coefficient pair. The DE technique used in the algorithm reduced significantly the embedding effects of DWT-base concealing technique and equally resolved the underflow/overflow issues encountered in concealing messages. More so, the scheme increased the system performances in areas of distortion tolerance and robustness to compression, salt and pepper, Gaussian noise, and cropping attacks when compared with other existing techniques. Nevertheless, the subjection of the system to a compression quality factor of 95% will destroy the embedded information. In other to enhance the security of digital image steganography Muhammad *et al.* (2017) introduced a two-level encrypted stego-key for protecting multilevel encrypted sensitive information. The stego-keys were produced by a pseudo-random number generator using a random permutation matrix, after which both bit xor and shuffling operations were performed for encrypting the output secret-keys. The system was further enhanced by encrypting the sensitive information using bit xor operation, blocks division key-based shuffling, with an encrypted stego-key. The cipher sensitive information, were lastly hiding in a predefined cover-mage region using least significant bits substitution methods. For successfully extraction of the sensitive information, the stego-key is required to perform a reverse of the embedding operation.

*Corresponding author: Idakwo, M. A. ✉ mondabutu@gmail.com ✉ Department of Computer Engineering, Faculty of Engineering, Ahmadu Bello University, Zaria.* 

In the hybrid cloud (Zhang *et al.*, 2018), Manoj *et al.* (2017) introduced a scalable and secured modern health record sharing with priority on maintaining patient information privacy. The system was implemented on a hybrid cloud with two subsystem domain (private and general domain) using information access requirements. The general domain users were based on professionalism while the private domain users' are the relatives and their patients. A stego-key of 128 bits length using key policy was used to enhance the system security. The stego-keys were based on RSA encryption standard while the electronic medical file was broken down into partition and encrypted using the AES encryption technique before storing it in the cloud. The encryption time and average system responses when performing HTTP requests for various user groups in both private and general domains were evaluated. The increase in sizes equally increases the AES encryption time which subjected the system to utilize extra bandwidth for transmission, longer seeking time and extra storage spaces. Elhoseny *et al.* (2018) Likewise, presented a secured steganography method for medical information exchange on the internet using a combination of AES and RSA encryption techniques. The information conveying image was transformed and decomposed into DWT sub-bands. In the encoding process, the secret information was converted into binary bits using ASCII and separated into odd and even parts. The RSA used public keys in encrypting the even parts while the AES encrypted the odd's part. The encrypted information size was reduced by compression before been inserted inside the cover-image with small effects on image quality. Although RSA encryption is computationally expensive when compared with AES, using the two encryption algorithm increases the computation complexity of the system and equally reduces the embeddable regions. Emad *et al.* (2018) similarly, presented a secured steganography algorithm for concealing secret text in digital images using LSB and IWT. The technique experimented with both colored and grayscale images as cover images. The secret information was concealed in the LSB of the approximation coefficients of the IWT. For effective assessment of the cover-image (grayscale and color) maximum embeddable regions, two variances of IWT- LSB scheme was developed. For the grayscale cover-images, the approximation coefficients LSB were substituted with the sensitive information whereas the color images used the approximation coefficients of the primary colors to conceal both the sensitive information and the sender's signature. The extraction phase of both techniques applied a median filter to discard noise from the stego-image. The results presented indicated that the system has the capacity of embedding up to 24,576 and 8192 in color and grayscale images respectively.

Walia *et al.* (2018) developed a new LSB substitution scheme using stego-key direction for the purpose of ensuring security on the internet. The system payload and imperceptibility were models as an optimization search problem. A LUDO base scanning technique was used in searching for cover-images free spaces. More so, cuckoo search optimization was integrated to select the optimal stego-key generated using Levy flight. The stego-key was made up of a shuffled 30 static bits and 60 bit optimally generated from the cuckoo search. The 90bits large keyspace is to increase the system resistance against brute force attacks. Furthermore, the key was used in shuffling sensitive information against cryptanalysis attacks. Finally, the sensitive information was randomly embedded in the cover-image using a chaotic map. Retrieving the sensitive information entails a reverse of all the operations. The technique obtained 44.09db and 0.97 as PSNR and SSIM respectively. However, encrypting a thoroughly shuffled sensitive information using a generated 90bits stego-key after performing a thorough search for all embeddable region increased the system computationally time.

*Corresponding author: Idakwo, M. A. ✉ mondabutu@gmail.com ✉ Department of Computer Engineering, Faculty of Engineering, Ahmadu Bello University, Zaria.* © *2020. Faculty of Technology Education, ATBU Bauchi. All rights reserved*

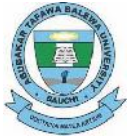

Xu and Nie (2018) presented an identity-based in time domain technique that allowed multiple sensitive information to be embedded in a cover-image for various users using fuzzy identity-based encryption. Several attributes were assigned to the various messages and the extraction process can only extract the right information with the right attribute. The system enabled performing hierarchical retrieving of information in steganography but nevertheless embedding both attributes and information reduces the system capable of holding information.

Valandar *et al.* (2019) presented a 3D sine chaotic map and IWT steganographic scheme. The IWT decomposed the colored image into four sub-bands. Each of the sub-band was divided into a non-overlapping 16 by 16 blocks acts as the embeddable regions for the sensitive information. The chaotic maps determine the sensitive information regions. Kalita *et al.* (2019) likewise, extends the regions for embedding information from one sub-band to three sub-bands excluding the approximation sub-band. A coefficient thresholds technique determines the sensitive information location. The coefficient LSB was directly replaced with the sensitive information bits to achieve imperceptibility and robustness. Nevertheless, it is difficult to obtain a threshold that is suitable for all images owing to images varying sizes, a variation of non-useful bit, quality and various sub-band sizes. Therefore may not be applicable for a real-time practical application having random images To achieve a high payload, Kadhim *et al.*, 2020 adopted an indirect concealing technique in the dual-tree complex wavelet transform with the aim of achieving a high payload with lesser distortion on the stego-image. A K-nearest neighbor machine learning model was adopted for classifying favorable images with expected high payload as cover-image. The images were classified as either textured or smooth using a 100 dataset for the training. The secret information was concealed in the dual-tree complex wavelet coefficient. The experimented result indicated that the system achieved a PSNR 50.55dB as its highest and an MSE of 0.1810 with a proposed security scheme to mitigate steganalyst attacks. Achieving a high payload is not solely on the technique employ for evaluating the redundant bits but equally, on the efficiency of the transform domain employ to transform the stego-image into its coefficient. This is because, a real-time application deals with a variance of images with different compression techniques, variation in light intensity, environmental factors, texture, noise from acquisition source amongst others. Thus this requires to increase the machine learning dataset in order to learn all the likely images that may be employed for the steganography system which may not be feasible.

From the literature surveyed, it is eminent that various steganographic techniques have been deployed for concealing secret information. With research priority focusing on payload, imperceptibility, security enhancement lower computational time. In spite of the time domain advantages in the guarantying of higher payload with lower computational time, their susceptibility to attacks restricts their uses in secured applications. Equally, the transform domain is robust to attacks, lower distortion tolerance with higher computationally time but their lower payload limits them to only low payload applications. Fortunately, hybridizing the time and frequency domain advantages may offer a good trade-off in achieving a high payload system that is immune to image processing attacks, distortion tolerance and is not computationally expensive.

*Corresponding author: Idakwo, M. A. ✉ mondabutu@gmail.com ✉ Department of Computer Engineering, Faculty of Engineering, Ahmadu Bello University, Zaria.* *© 2020. Faculty of Technology Education, ATBU Bauchi. All rights reserved*

## CONCLUSION

Steganography has evolved from the simplest method of replacing the least pixel bits with the sensitive information to highly sophisticated techniques which use the frequency and hybrid domain owing to obvious advantages of imperceptibility and robustness to attacks. However, the pertinent problems that have span to date have been on how to maximize embeddable regions to accommodate more information amidst limited redundancy bits while maintaining imperceptibility, resistances to attacks and lower computational time. Solving these salient issues we lead to a wider application of steganography in real-time applications with higher privacy demands.

*Corresponding author: Idakwo, M. A. ✉ mondabutu@gmail.com ✉ Department of Computer Engineering, Faculty of Engineering, Ahmadu Bello University, Zaria.* 

*Corresponding author: Idakwo, M. A. ✉ mondabutu@gmail.com ✉ Department of Computer Engineering, Faculty of Engineering, Ahmadu Bello University, Zaria.* © 2020. Faculty of Technology Education, ATBU Bauchi. All rights reserved

*Corresponding author: Idakwo, M. A. ✉ mondabutu@gmail.com ✉ Department of Computer Engineering, Faculty of Engineering, Ahmadu Bello University, Zaria.* © 2020. Faculty of Technology Education, ATBU Bauchi. All rights reserved